\documentclass[12pt]{article}

\newcommand{\p}{\prime}
\newcommand{\q}{\quad}
\newcommand{\mR}{{\bf R}} 
\newcommand{\be}{\begin{equation}}
\newcommand{\ee}{\end{equation}}
\newcommand{\bea}{\begin{eqnarray}}
\newcommand{\eea}{\end{eqnarray}}

\begin{document}
\begin{titlepage}
 
 
\vspace{1in}
 
\begin{center}
\Large
{\bf Dynamical Systems in Cosmology}
 
\vspace{1in}

\normalsize

Alan A. Coley

\normalsize
\vspace{.7in}

 {\em Department of Mathematics and Statistics, \\ Dalhousie University,
Halifax, NS, B3H 3J5,
Canada} \\

\end{center}
 
\vspace{.5in}
 
\baselineskip=24pt
\begin{abstract}
\noindent
Dynamical systems theory is especially well-suited for determining the possible 
asymptotic states (at both early and late times) of cosmological models, 
particularly when the governing equations are a finite system of 
autonomous ordinary differential equations. We begin with a brief review of dynamical systems theory.
We then discuss cosmological models as dynamical systems and point out the
important role of self-similar models. We review the asymptotic properties of 
spatially homogeneous perfect fluid models in general relativity. We then discuss
some results concerning scalar field models with an exponential potential
(both with and without barotropic matter). Finally, we discuss some isotropic
cosmological models derived from the string effective action.
\end{abstract}

\vspace{.7in}
\noindent Electronic mail: aac@mscs.dal.ca

\end{titlepage}

\section{Introduction}

The governing equations of the most commonly studied cosmological models 
are a system of autonomous ordinary
differential equations (ODEs). Since our main goal is to give a qualitative 
description
of these models, a dynamical systems approach is undertaken. Usually, 
a dimensionless (logarithmic) time variable, $\tau$, is introduced
so that the models are valid for all times (i.e., $\tau$ assumes all 
real values).
A normalised set of variables are then chosen
for a number of reasons. First, this normally leads to a 
compact dynamical system. Second, these variables
are well-behaved and often have a direct physical interpretation. 
Third, due to a symmetry in the equations, one of the equations 
decouple 
(in general relativity the expansion is used to normalize the 
variables in ever expanding models
whence the Raychaudhuri equation decouples) and
the resulting simplified reduced system is then studied.
The singular points of the reduced system then correspond to 
dynamically evolving self-similar cosmological models.
More precisely, using the dimensionless time variable and a 
normalised set of variables, the governing ODEs define a flow and  the evolution
of the cosmological models can then be analysed by studying 
the orbits of this flow
in the physical state space, which is a subset of Euclidean space.
When the state space is compact, each orbit will have a non-empty 
$\alpha$-limit set and 
$\omega$-limit set, and hence there will be a both a past attractor 
and a future attractor in the state
space.

\subsection{Self-similarity} 

Self-similar 
solutions of the Einstein field equations (EFE) play an important role 
in describing the asymptotic properties of more general models. 
The energy-momentum tensor of a perfect fluid given by
\be
 T_{ab} = (\mu + p) u_a u_b + pg_{ab},  
 \ee
where $u^a$ is the normalized fluid $4$-velocity, $\mu$ is the density 
and $p$ is the pressure satisfying a linear barotropic equation of 
state of the form
\be
p = (\gamma - 1) \mu,  
\ee
where $\gamma$ is a constant. 
The existence of 
a self-similarity  of the {\it first kind} can be 
invariantly 
formulated in terms of the existence of a {\it homothetic vector} \cite{Cahill}.  
For a general spacetime a proper homothetic
vector (HV) is a vector field $\mbox{\boldmath $\xi$}$ which satisfies  
\be
{\cal L}_\xi g_{\mu v} = 2g_{\mu v}, 
\ee
where $g_{\mu v}$ is the metric and ${\cal L}$ denotes Lie differentiation along 
$\mbox{\boldmath $\xi$}$.  An arbitrary constant on the right-hand-side of (1.3) 
has been rescaled to unity.  If this constant is
zero, i.e., ${\cal L}_\xi g_{u v}=0$, then $\mbox{\boldmath $\xi$}$ is a 
Killing vector.  A homothetic motion 
or homothety captures the geometric notion of ``invariance under scale 
transformations''.

Self-similar models are often 
related to the asymptotic states of more general models \cite{hsu}. 
In particular, self-similar
models play an important role in the asymptotic
properties of spatially homogeneous models, spherically
symmetric models, $G_2$ models and silent universe
models  \cite{bruni}. We will focus on
spatially homogeneous models here  \cite{EM,kramer} 
We note that the self-similar 
Bianchi models of relevence below are {\it transitively} self-similar (in the sense
that the
orbits of the $H_4$ are the whole spacetime). 
Self-similar spherically symmetric models have been studied by many 
authors and have been recently reviewed in Carr and Coley \cite{cc}.
$G_2$ models, which contain two commuting spacelike KV
acting orthogonally transitively,
 have been discussed in Wainwright and Ellis  \cite{WE} (hereafter denoted WE).
Other exact 
homothetic models, including for example, plane-symmetric models
have been discussed in Kramer et al.  \cite{kramer} and  Carr and Coley  \cite{cc}.

{\em Exact solutions}

Let us review some exact self-similar solutions that are of particular importance.

${\bullet}$ Minkowski space ($M$):
\begin{eqnarray}
ds^2 & = & -dt^2 + dx^2 + dy^2 + dz^2  \\
\mbox{\boldmath $\xi$} & = & t \partial_t + x\frac{\partial}{\partial x} + 
y \frac{\partial}{\partial y
}
+ z \frac{\partial}{\partial z}. 
\end{eqnarray}

In addition to flat Minkowski spacetime, 
all $(k = 0, \pm 1)$ Friedmann--Robertson--Walker (FRW) models admit a timelike HV
in the special case of stiff matter $(\gamma =2)$.  Otherwise, 
only the $k = 0$ models admit a HV, and this occurs for all such
models in which  $p = (\gamma -1) \mu$ and hence the scale function has 
power-law depend
on time.

${\bullet}$ $k = 0$, FRW $(F_0)$:
\begin{eqnarray}
ds^2 & = & -dt^2 + t^{ 4/3 \gamma} (dx^2 + dy^2 + dz^2)  \\
\mbox{\boldmath $\xi$} & =&  t \partial_t + \left(1 - \frac{2}{3 \gamma} \right)
\left( x \frac{2}{\partial x} + y \frac{\partial}{\partial y} + z 
\frac{\partial}{\partial z}\right) 
\end{eqnarray}
where
\be
p = (\gamma-1) \mu; \enskip \mu = \frac{4}{3 \gamma^2} t^{-2} \enskip \left(
\mbox{\boldmath $u$}  = 
\frac{\partial}{\partial t} \right).  
\ee

All transitively self-similar orthogonal spatially homogeneous 
perfect fluid solutions (with $\frac{2}{3} < \gamma \leq 2$) 
and spatially homogeneous vacuum solutions are summarized in
Hsu and Wainwright  \cite{hsu}. In particular, the Kasner vacuum solution is self-similar.

$\bullet$ Kasner (vacuum) ($K$):
\be
ds^2  = -dt^2 + t^{2 p_1} dx^2 + t^{2 p_2} dy^2 + t^{2p_3}dz^2  
\ee
with
\begin{eqnarray}
\Sigma p_i & = &\Sigma p^2_i = 1  \\
\mbox{\boldmath $\xi$} & = & t \partial_t + (1 - p_1) x \frac{\partial}{\partial x} 
+ (1 - p_2) y\frac{\partial}{\partial y} + (1 - p_3) z\frac{\partial}
{\partial z}. 
c
\end{eqnarray}

\subsection{Brief Survey of Techniques in Dynamical Systems}

This section will briefly review some of the results of dynamical 
systems theory.

\noindent
{\bf Definition 1}
A {\it singular point} of a system of autonomous ODEs 
\begin{equation} \dot{x} = f(x)\label{DE}\end{equation}
 is a point $\bar{x} \in {\mbox{\boldmath R}}^n$ such that $f(\bar{x})=0$.

\noindent
{\bf Definition 2}
Let $\bar{x}$ be a singular point of the DE (\ref{DE}).  The point $\bar{x}$ is 
called a {\it hyperbolic} singular point if $Re(\lambda_i) \neq 0$ for all 
eigenvalues, $\lambda_i$, of the Jacobian of the vector field $f(x)$ evaluated 
at $\bar{x}$.  Otherwise the point is called {\it non-hyperbolic}.

\noindent
{\bf Definition 3}
Let $x(t)=\phi_a(t)$ be a solution of the DE (\ref{DE}) with initial condition 
$x(0)=a$.  The flow $\{g^t\}$ is defined in terms of the solution function 
$\phi_a(t)$ of the DE by
\begin{center}
$g^ta=\phi_a(t).$
\end{center}

\noindent
{\bf Definition 4} The orbit through $a$, denoted by $\gamma(a)$ is defined by
\begin{center}
$\gamma(a)=\{x\in \mR^n | x=g^ta, $ for all $t \in \mR\}$.
\end{center}

\noindent
{\bf Definition 5} Given a DE (\ref{DE}) in \mR$^n$, a set 
$S \subseteq \mR^n$ is called an invariant set for the DE 
if for any point $a \in S$ the orbit through $a$ lies entirely 
in $S$, that is $\gamma(a) \subseteq S$.

\noindent
{\bf Definition 6}  Given a DE (\ref{DE}) in \mR$^n$, with flow 
$\{g^t\}$, a subset $S \subseteq \mR^n$ is said to be a trapping 
set of the DE if it satisfies:
1.   $S$ is a closed and bounded set,
2.   $a \in S$ implies that $g^ta \in S$ for all $t \geq 0$.

Qualitative analysis of a system begins with the location of singular
points.  Once the singular points of a system of ODEs are obtained, it
is of interest to consider the dynamics in a local neighbourhood of
each of the points.  Assuming that the vector field $f(x)$ is of class
$C^1$ the process of determining the local behaviour is based on the
linear approximation of the vector field in the local neighbourhood of
the singular point $\bar{x}$.  In this neighbourhood
\be
{f(x) \approx Df(\bar{x})(x-\bar{x})}\label{linapp}
\ee
where $Df(\bar{x})$ is the Jacobian of the vector field at the singular 
point $\bar{x}$.  The system (\ref{linapp}) is referred to as the 
{\it linearization of the DE at the singular point}.
Each of the singular points can then be classified according to the 
eigenvalues of the Jacobian of the linearized vector field at the point.

The classification then follows from the fact that if the singular point 
is hyperbolic in nature the flows of the non-linear system and it's linear 
approximation are {\it topologically equivalent} in a neighbourhood of the 
singular point.  This result is given in the form of the following theorem:

{\bf Theorem 1:  Hartman-Grobman Theorem}
Consider a DE: $\dot{x}=f(x)$, where the vector field $f$ is of class $C^1$.  
If $\bar{x}$ is a hyperbolic singular point of the DE then there exists a 
neighbourhood of $\bar{x}$ on which the flow is topologically equivalent to 
the flow of the linearization of the DE at $\bar{x}$.

Given a linear system of ODEs: 
\be{\dot{x}=Ax,}\label{linsys}\ee
where A is a matrix with constant coefficients, it is a straightforward matter to 
show that if the eigenvalues of the matrix A are all positive the solutions 
in the neighbourhood of $\bar{x}=0$ all diverge from that point.  This point 
is then referred to as a source.  Similarly, if the eigenvalues all have negative 
real parts all solutions converge to the singular point $\bar{x}=0$, and the 
point is referred to as a sink.  Therefore, it follows from topological 
equivalence that if all eigenvalues of the Jacobian of the vector field 
for a non-linear system of ODEs have positive real parts the point is 
classified as a source (and all orbits diverge from the singular point), 
and if the eigenvalues all have negative real parts the point is classified 
as a sink.

In most cases the eigenvalues of the linearized system (\ref{linapp}) will 
have eigenvalues with both positive, negative and/or zero real parts.  In 
these cases it is important to identify which orbits are attracted to the 
singular point, and which are repelled away as the independent variable 
(usually $t$) tends to infinity.

For a linear system of ODEs, (\ref{linsys}), the phase space \mR$^n$ is spanned 
by the eigenvectors of $A$.  These eigenvectors divide the phase space into 
three distinct subspaces; namely:
\begin{center}
\begin{tabular}{ll}
The {\it stable subspace}     & $E^s=$ span$(s_1, s_2, ... s_{ns})$ \\
The {\it unstable subspace}   & $E^u=$ span$(u_1, u_2, ... u_{nu})$\\
The {\it centre subspace}     & $E^c=$ span$(c_1, c_2, ... c_{nc})$ 
\end{tabular}
\end{center}
where $s_i$ are the eigenvectors who's associated eigenvalues have negative 
real part, $u_i$ those who's eigenvalues have positive real part, and $c_i$ 
those who's eigenvalues have zero eigenvalues.  Flows (or orbits) in the stable 
subspace asymptote in the future to the singular point, and those in the 
unstable subspace asymptote in the past to the singular point.

In the non-linear case, the topological equivalence of flows allows 
for a similar classification of the singular points.  The equivalence 
only applies in directions where the eigenvalue has non-zero real parts.  
In these directions, since the flows are topologically equivalent, there 
is a flow {\it tangent} to the eigenvectors.  The phase space is again 
divided into stable and unstable subspaces (as well as centre subspaces).  
The {\it stable manifold} $W^s$ of a singular point is a differential 
manifold which is tangent to the stable subspace of the linearized 
system ($E^s$).  Similarly, the {\it unstable manifold} is a differential 
manifold which is tangent to the unstable subspace ($E^u$) at the singular 
point.  The centre manifold, $W^c$, is a differential manifold which is 
tangent to the centre subspace $E^c$.  It is important to note, however, 
that unlike the case of a linear system, this centre manifold, $W^c$ will 
contain all those dynamics not classified by linearization (i.e., the 
non-hyperbolic directions).  In particular, this manifold may contain regions 
which are stable, unstable or neutral.  The classification of the dynamics in 
this manifold can only be determined by utilizing more sophisticated methods, 
such as centre manifold theorems or the theory of normal forms (see \cite{Wiggins1990a}).

Unlike a linear system of ODEs, a non-linear system allows for
singular structures which are more complicated than that of the
singular points, fixed lines or periodic orbits.  These structures
include, though are not limited to, such things as heteroclinic and/or homoclinic orbits
and non-linear invariant sub-manifolds (for
definitions see \cite{Wiggins1990a}).  Sets of non-isolated
singular points often occur in cosmology and therefore their stability will be examined more rigorously.

{\bf Definition 7}:  A set of non-isolated singular points is said to be normally 
hyperbolic if the only eigenvalues with zero real parts are those whose 
corresponding eigenvectors are tangent to the set.

Since by definition any point on a set of non-isolated singular points will 
have at least one eigenvalue which is zero, all points in the set are 
{\it non-hyperbolic}.  A set which is normally hyperbolic can, however, 
be completely classified as per it's stability by considering the signs 
of the eigenvalues in the remaining directions  (i.e.,for a curve, in the 
remaining $n-1$ directions) \cite{Aulbach1984a}.

The local dynamics of a singular point may depend on one or more 
arbitrary parameters.  
When small continuous
changes in the parameter result in dramatic
changes in the dynamics, the singular point is said to undergo a 
{\em bifurcation}\label{bifurcation}.
The values of the parameter(s) which result in a bifurcation at the singular
point can often be located by examining the linearized system.  Singular point
bifurcations will only occur if one (or more) of the eigenvalues of the 
linearized systems are a function of the parameter.  The bifurcations
are located at the parameter values for which the real part of an eigenvalue
is zero. 

The future and past asymptotic states of a non-linear system 
may be represented by any singular or periodic structure.
In the case of a plane system (i.e., in two-dimension phase 
space), the possible asymptotic states can be given explicitly.  This result 
is due to the limited degrees of freedom in the space, and the fact that the 
flows (or orbits) in any dimensional space cannot cross.  The result is given 
in the form of the following theorem:

{\bf Theorem 2: Poincare-Bendixon Theorem}:  Consider the system of ODEs 
$\dot{x}=f(x)$ on \mR$^2$, with $f \in C^2$, and suppose that there 
are at most a finite number of singular points (i.e., no non-isolated 
singular points).  Then any compact asymptotic set is one of the following; 
1. a singular point, \enskip
2. a periodic orbit, \enskip
3. the union of singular points and heteroclinic or homoclinic orbits.

This theorem has a very important consequence in that if the existence of 
a closed (i.e., periodic, heteroclinic or homo-clinic) orbit can be ruled 
out it follows that all asymptotic behaviour is located at a singular point.

The existence of a closed orbit can be ruled out by many methods, the most 
common is to use a consequence of Green's Theorem, as follows:

{\bf Theorem 3: Dulac's Criterion}:  If $D \subseteq R^2$ is a simply 
connected open set and 
$\nabla(Bf) = \frac{\partial}{\partial x_1}
	(Bf_1)+\frac{\partial}{\partial x_2}(Bf_2) > 0, $ or $(<0)$ 
for all $x \in D$ where $B$ is a $C^1$ function, then the DE $\dot{x}=f(x)$ 
where $f \in C^1$ has no periodic (or closed) orbit which is contained in $D$.

A fundamental criteria of the Poincare-Bendixon theorem is that the phase 
space is two-dimensional.  When the phase space is of a higher dimension 
the requirement that orbits cannot cross does not result in such a decisive 
conclusion.  The behaviour in such higher-dimensional spaces is known to be 
highly complicated, with the possibility of including such phenomena as 
recurrence and strange attractors  \cite{Guckenheimer1983a} .
  For that reason, the analysis of non-linear systems in 
spaces of three or more dimensions cannot in general progress much further 
than the local analysis of the singular points (or non-isolated singular sets).  
The one tool which does allow for some progress in the analysis of higher 
dimensional systems is the possible existence of monotonic functions.  

{\bf Theorem 4: LaSalle Invariance Principle}:  Consider a DE 
$\dot{x}=f(x)$ on \mR$^n$.  Let $S$ be a closed, bounded and 
positively invariant set of the flow, and let $Z$ be a $C^1$ monotonic 
function.  Then for all $x_0\in S$, 
\begin{center}
$w(x_0) \subset \{x \in S | \dot{Z}=0\}$
\end{center}
where $w(x_0)$ is the forward asymptotic states for the orbit with 
initial value $x_0$; i.e., a $w$-limit set \cite{Tavakol1997a}.

\noindent
This principle has been generalized to the following result:

{\bf Theorem 5: Monotonicity Principle} (see \cite{LeBlanc1997a}).  
Let $\phi_t$ be a flow on \mR$^n$ with $S$ an invariant set.  
Let $Z : S \rightarrow \mR$ be a $C^1$ function whose range 
is the interval $(a,b)$, where $a \in \mR\cup \{-\infty\}$, 
$b \in \mR \cup \{\infty\}$ and $a<b$.  If Z is decreasing 
on orbits in $S$, then for all $X \in S$,
\begin{center}
$\omega(x) \subseteq \{s \in \bar{S}\setminus S | lim_{y\rightarrow s} Z(y) \neq b\},$ \\
$\alpha(x) \subseteq \{s \in \bar{S}\setminus S | lim_{y\rightarrow s} Z(y) \neq a\},$
\end{center}
where $\omega(x)$ and $\alpha(x)$ are the forward and backward limit set of 
$x$, respectively (i.e., the $w$ and $\alpha$ limit sets.)

\section{Spatially Homogeneous Perfect Fluid Models}
\setcounter{equation}{0}

Many people have studied self-similar spatially homogeneous models, 
both as exact solutions and in the context of 
qualitative analyses (see WE and Coley \cite{Coley1997b} and references therein).
Exact spatially homogeneous solutions were first displayed in 
early papers  \cite{Eardley1974}; however, 
it was not until after 1985 that many of them were recognized by Wainwright \cite{Wainwright1985} 
and Rosquist and Jantzen  \cite{J1984,R1984} as being self-similar.  The complete set of  
 self-similar orthogonal spatially
homogeneous perfect fluid and vacuum solutions were  given   
by Hsu and Wainwright  \cite{hsu} and they have also been reviewed in WE.  
Kantowski-Sachs models were studied by Collins \cite{Collins1977}.   

Spatially homogeneous models have attracted considerable
attention since the governing equations reduce to a relatively simple 
finite-dimensional
dynamical system, thereby enabling the models
to be studied by standard qualitative techniques.  Planar systems were 
initially analyzed by
Collins  \cite{Collins1971, Collins1974} and a comprehensive study of general
Bianchi models was made by Bogoyavlenski and Novikov  \cite{BN} and
Bogoyavlenski  \cite{bog} and more recently (using automorphism variables and
Hamiltonian techniques) by Jantzen and Rosquist \cite{J1984, R1984,
RJ1988, RUJ}.
Perhaps the most illuminating approach has been that of
Wainwright and collaborators  \cite{hsu, whsu, HW93}, 
in which the more physically or geometrically natural 
expansion-normalized (dimensionless) configuration variables are 
used. In this case, the
physically admissible states typically lie within a bounded region,
the dynamical system remains analytic both in the physical
region and its boundaries, and the asymptotic
states typically lie on the boundary represented by exact physical
solutions rather than having singular behaviour.
We note that  the 
physically admissible states do not lie in a bounded region for Bianchi 
models of types VII$_0$, VIII and IX;
see WE for details.

Wainwright utilizes the orthonormal frame method
 \cite{EM} and introduces expansion-normalized
(commutation function) variables and a new ``dimensionless'' time 
variable to study spatially
homogeneous perfect fluid models satisfying $p = (\gamma - 1) \mu$.    The
equations governing the models form an $N$-dimensional
system of coupled autonomous ODEs.  When the ODEs
are written in expansion-normalized
variables, they admit a symmetry which allows the equation for the
time evolution of the expansion $\theta$  (the Raychaudhuri equation) to decouple.  
The reduced $N-1$-dimensional dynamical system is then
studied.  At all of the singular points of the reduced
system,  $\dot{\theta}$ is proportional to $\theta^2$  and hence all such
points correspond to transitively self-similar cosmological models  \cite{hsu}.  
This is why the   self-similar models play an important 
role in describing the asymptotic dynamics of the Bianchi models.
For orthogonal Bianchi models of class A, the resulting reduced 
state space is five-dimensional  \cite{whsu}. Orthogonal Bianchi
cosmologies of class B were studied
by Hewitt and Wainwright  \cite{HW93} and are governed by
a five-dimensional system of analytic ODEs with constraints.

Two perfect-fluid models were studied by Coley and Wainwright  \cite{CW}.
In further work, imperfect fluid Bianchi models were
studied under the assumption that all physical quantities
satisfy ``dimensionless equations of state'', thereby 
ensuring that the singular points of the resulting reduced
dynamical system are represented by exact self-similar solutions.  Models satisfying the linear
Eckart theory of irreversible thermodynamics were studied by Burd and Coley  \cite{BurdColey1994} 
 and Coley and van den Hoogen  \cite{CvdH94}, those satisfying the truncated 
causal theory 
of Israel-Stewart by Coley and van den Hoogen \cite{CvdH95},
and those satisfying the full (i.e., non-truncated) relativistic 
Israel-Stewart 
theory by Coley et al.  \cite{CHM}.  Self-similar solutions also 
play an important role in
describing the dynamical behaviour of cosmological models 
close to the Planck time in general
relativity with scalar fields
\cite{CLW,BCHIO}, in scalar-tensor theories
of gravity  \cite{Coley99}, and particularly in the low-energy
limit in supergravity theories from string theory and other higher-dimensional
gravity theories.

\subsection{Some Simple Examples}

The expansion scalar $\theta$, shear scalar $\sigma \enskip (\sigma^2 = \frac{1}{2} 
\sigma_{ab} \sigma^{ab}$), and Ricci
$3$-curvature (orthogonal to $u^a$) $^3R$, are defined in Ellis \cite{Ellis1971}.

\noindent
{\it Basic equations:}
\begin{eqnarray}
&&  \mbox{Raychaudhuri eqn.}: \dot{\theta}  =   - \frac{1}{3} \theta^2 - 
2 \sigma^2 - \frac{1}{3} (\mu + 3p).  \\
 && \mbox{Conservation eqn.} : \dot{\mu}   =   - (\mu + p) \theta.   \\
&&  \mbox{Generalized Friedmann eqn.}: \theta^2   =  
   - \frac{1}{3}\, ^3\!R + 3 \sigma^2 + 3 \
mu.  \\
 && \mbox{Ricci identity: for } \dot{\sigma}  \qquad \mbox{\cite{Ellis1971}}.   
\end{eqnarray}
We define the expansion-normalized variables ($\theta > 0)$:
\begin{eqnarray} 
\Omega & =& \frac{3 \mu}{\theta^2}, \qquad \quad\mbox{density parameter}  \\
\beta & =& 2 \sqrt{3} \frac{\sigma}{\theta}, \quad \enskip  
\mbox{normalized rate of shear} \\
\frac{d \tau}{d t}& =& \frac{1}{3} \theta, \quad  \qquad 
[\tau \rightarrow -\infty \mbox{
 as } 
t \rightarrow 0^+]   
\end{eqnarray}
A dash (dot) denotes  differentiation
with respect to $\tau \, (t)$.  The equation of state is 
given by Eq. (1.2),
where $1 \leq \gamma \leq 2$ 
for normal matter
and models with $0 \leq \gamma < \frac{2}{3}$ are of interest in connection
with inflationary models of the universe (see, for example, Wald \cite{Wald83}).
The weak energy condition implies that
$$\mu \geq 0. $$
\noindent
{\bf A. FRW}: The metric is given by
\be
ds^2 = -dt^2 + R^2(t) \left[ \frac{dr^2}{1-kr^2} + r^2 d\Omega^2 \right], 
\ee 
where $k$ is the curvature constant.  Here $\sigma = 0$ ($\theta = 3\dot{R}/R)$, 
and the equations reduce to the single ODE:
\be
\Omega' = (2 - \gamma) \Omega (1-\Omega), 
\ee
where $\Omega \geq 0$ and $1 - \Omega = -\frac{3}{2} \, ^3\!R \theta^{-2}$, 
so that $\Omega < 1, = 1, > 0$ according to whether
$k < 0, = 0, > 0$ (models open, flat or closed), respectively.

\noindent
$\bullet$ Singular points $(\gamma \neq 2)$: 

$\begin{array}{ll   l}
\bullet \enskip \Omega =1 \enskip (F_0); & \mbox{past attractor} &
 \mbox{(flat FRW model describes the dynamics  }\\
   &     & \mbox{near the big bang)}.\\
&&\\
\bullet \enskip \Omega = 0 \enskip (M); & \mbox{future attractor} & \mbox{(open models approach Milne form of} \\
& & \mbox{Minkowski space at later times).}  
\end{array}$

\noindent
[For closed models $\theta = 0$ at the point of maximum expansion,
so the models are only valid up to this point. The models recollapse and 
asymptotically approach the flat FRW model $F_0$ to the future].

\noindent
{\bf B. Bianchi V:}  The metric is given by 
 \be
 ds^2 = -dt^2 + A^2 (t) dx^2 + e^{2x}(B^2(t) dy^2 + A^4 (t) B^{-2} (t) dz^2), 
\ee 
 where $\theta = 3 \frac{\dot A}{A}$, $\sigma^2 = \left( \frac{\dot A}{A} - \frac{\dot B}
{B}\right)$, and $^3R = -\frac{6}{A^2}$.  The Ricci identity leads to 
\be
\dot{\sigma} = - \theta \sigma. 
\ee
The EFE then reduce to the plane autonomous system:
\begin{eqnarray}
\Omega' & = & (2 - 3 \gamma) \Omega (1 - \Omega) +  \beta^2 
\Omega \nonumber\\
\beta' & = & \frac{\beta}{2}[(3 \gamma - 2) \Omega + \beta^2 -4],  
\end{eqnarray} 
where 
$$\beta^2 + 4 \Omega - 4 = - \frac{1}{A^2 \theta^2} \leq 0 $$
(from the generalized Friedmann eqn.) and
$$\Omega \geq 0, $$
so that phase space is compact.
 
$\begin{array}{l l  l }
\bullet \enskip \mbox{ invariant sets:} &\beta = 0  \mbox{       (FRW, dealt with above)}.\\
&  \\
&\beta^2 + 4 \Omega - 4 = 0 \mbox{ (Bianchi I, boundary)} & 
\end{array}$

In the Bianchi I (boundary) case we have 
a single  ODE and the models evolve from  $K$ to  $F_0$.
In  addition, the equations above are symmetric about the $\Omega$-axis.

$\bullet$
   singular points $(\Omega, \beta)$  [$\gamma \neq 2$]:

$\begin{array}{ l  l l  l}  
\quad  \bullet &  (0, 0) \quad (M): &   \mbox{ future attractor} & \\
&&&\\
         & (1, 0) \quad (F_0): & \mbox{ saddle} & \mbox{(past attractor for open FRW models,}\\
&&& \mbox{intermediate asymptote otherwise).}\\
&&&\\ 
\quad \bullet &   \left. \begin{array}{ l  l} 
  (0, 2) & (K_+):\\
&\\
(0, -2) & (K_ -):   
\end{array} \right\} & \mbox{ past attractor} & \mbox{(initial Kasner cigar singularity).}
 \end{array}$
  
In the special case $\gamma = 2$, the boundary $\beta^2 + 4 \Omega - 4 = 0$ becomes a line of 
repelling singular points.  The corresponding 
Bianchi I stiff matter models
evolve from the general self-similar (stiff matter) Bianchi I solution of Jacobs \cite{Jacobs1968} 
to the Milne (flat) universe  \cite{Collins1971}.

\subsection{Asymptotic States of Bianchi Models}

We now discuss  
the asymptotic states of Bianchi models, again assuming the linear equation of state (1.2). 
We will summarize the work of Wainwright and Hsu  \cite{whsu}
and Hewitt and Wainwright  \cite{HW93}, who studied the asymptotic states of orthogonal spatially 
homogeneous models in terms of attractors of the 
associated dynamical system for class $A$ and class $B$ models, respectively. 
Due to the existence of monotone functions, it is known
that there are no periodic or recurrent orbits in class A models.
Although ``typical'' results can be proved in a number 
of Bianchi type B cases, these are not ``generic'' due to the lack of knowledge of appropriate monotone functions.
In particular, there are no sources or sinks in the Bianchi invariant sets 
$B^{\pm}_\alpha$ (VIII) or $B^\pm$(IX).

The key results are as follows:

\begin{enumerate} 
\item[$\bullet$] A large class of orthogonal spatially homogeneous models (including all class B models) are asymptotically self-similar
at the initial singularity and are approximated by exact perfect fluid or 
vacuum self-similar
power-law models.  Examples include
self-similar Kasner vacuum models or self-similar locally rotationally
symmetric (class III) Bianchi type II perfect fluid models  \cite{Collins1971, CollinsStewart1971, Doroshkevich1973}.
\end{enumerate}

\noindent
However, this behaviour is not generic; general orthogonal models of Bianchi types IX and VIII 
have an oscillatory behaviour with chaotic-like characteristics, 
with the matter density becoming dynamically negligible as one follows the evolution into the past towards the initial singularity. 
 Ma and Wainwright  \cite{MW1994} 
show that the orbits of the associated cosmological 
dynamical system are negatively asymptotic to a lower two--dimensional 
attractor. 
This is the union of three ellipsoids in ${\bf R}^5$ consisting 
of the Kasner ring joined by
Taub separatrices; the orbits spend most of the time near the 
Kasner vacuum 
singular points.  Clearly  the self-similar Kasner models play a 
primary role in the 
asymptotic behaviour of these models.

\begin{enumerate} 
\item[$\bullet$]  Exact self-similar power-law models can also 
approximate general Bianchi models at intermediate 
stages of their evolution  (e.g., radiation Bianchi VII$_h$ models  \cite{Doroshkevich1973}). 
 Of special interest are those models which can be 
approximated by an isotropic solution at an intermediate stage of their evolution (e.g., those models whose orbits spend a 
period of time near to a flat Friedmann singular point).
\end{enumerate}

\noindent
This last point is of particular importance in relating Bianchi models to the 
real Universe, and is discussed further in general terms in WE (see, especially, Chapter 15) 
and specifically for Bianchi VII$_h$ models in Wainwright et al.  \cite{WCEH}.  In particular, 
the flat Friedmann singular point is 
universal in that it is contained in the state space of each Bianchi type. 
Isotropic intermediate behaviour has also been found in tilted Bianchi V models
 \cite{HW1992}, and it appears that many tilted models have isotropic 
  intermediate behaviour (see WE).

\begin{enumerate} 
\item[$\bullet$]  Self-similar solutions can describe the behaviour 
of Bianchi models at late times (i.e., as $t \rightarrow \infty$). 
Examples include self-similar flat space and self-similar homogeneous 
vacuum plane waves  \cite{Collins1971, Wainwright1985}.
\end{enumerate}

\noindent
All models expand indefinitely except for the Bianchi type IX models.  The 
question of which Bianchi models can isotropize was 
addressed in the famous paper by Collins and Hawking \cite{CollinsHawking73}, in which it was shown that, 
for physically reasonable matter, the set of homogeneous initial data that give rise to models that isotropize asymptotically to the future is of zero 
measure in the space of all homogeneous initial data (see also WE).

\noindent
All vacuum models of Bianchi (B) types IV, V, VI$_h$ and 
VII$_h$ are asymptotic to plane wave states to the future.
Type V models tend to the Milne form of flat spacetime  \cite{HW93}. Typically, and perhaps generically \cite{HW93}, 
non-vacuum models are asymptotic in the future
to either plane-wave vacuum solutions  \cite{Doroshkevich1973} 
or non-vacuum Collins type VI$_h$ solutions  \cite{Collins1971}.

\noindent
Bianchi (A) models of types VII$_o$ (non-vacuum) and VIII 
expand indefinitely but are found to have oscillatory (though 
non-chaotic) behaviour in the Weyl curvature (see, for example,  \cite{Wainwright1999}).
Bianchi type IX models obey the ``closed universe recollapse'' conjecture
 \cite{LinWald1989}.  All orbits in the Bianchi invariant 
sets $B^\pm_\alpha(VII_0$) ($\Omega > 0), B^\pm_\alpha(VIII)$ and $B^\pm(IX)$ are 
positively departing; in order to analyse the
future asymptotic states of such models it is necessary to compactify phase-space.  The description of these models in terms
of conventional expansion-normalized variables is only valid up to the point of 
maximum expansion (where $\theta = 0)$.

\section{Scalar Field Models}
\setcounter{equation}{0}

\subsection{Background}

A variety of theories of fundamental physics predict the existence of
scalar fields \cite{Green1987a,Olive1990a,Billyard1999h}, motivating
the study of the dynamical properties of scalar fields in cosmology.
Indeed, scalar field cosmological models are of great importance in
the study of the early universe, particularly in the investigation of
inflation \cite{Olive1990a,Guth1981a,Linde1987a}.  Recently there has also been
great interest in the late-time evolution of scalar field models.
`Quintessential' scalar field models (or slowly decaying cosmological
constant models) \cite{Caldwell1998a,Bahcall1999a} give rise to a
residual scalar field which contributes to the present energy-density
of the universe that may alleviate the dark matter problem and can
predict an effective cosmological constant which is consistent with
observations of the present accelerated cosmic expansion
\cite{Perlmutter1999a,Riess1998a}.

Models with a self-interaction potential with an exponential
dependence on the scalar field, $\phi$, of the form
\be
V = \Lambda  e^{k \phi},
\ee
where $\Lambda$ and $k$ are positive constants, have been the subject
of much interest and arise naturally from theories of gravity
such as scalar-tensor theories or string theories
\cite{Billyard1999h}.  Recently, it has been argued that a scalar
field with an exponential potential is a strong candidate for dark
matter in spiral galaxies \cite{Guzman1998a} and is consistent with
observations of current accelerated expansion of the universe
\cite{Huterer1998a}.

A number of authors have studied scalar field cosmological models with
an exponential potential within general relativity.  Homogeneous and isotropic
FRW models were studied by Halliwell
\cite{Halliwell1987a} using phase-plane methods.
Homogeneous but anisotropic models of Bianchi types I and III (and
Kantowski-Sachs models) were studied by Burd and Barrow
\cite{Burd1988a}, Bianchi type
I models were studied by Lidsey \cite{Lidsey1992a} and Aguirregabiria {\em et
al.} \cite{Aguirregabiria1993b}, and Bianchi models of types III and VI were
studied by Feinstein and Ib\'{a}\~{n}ez \cite{Feinstein1993a}.
A qualitative analysis of Bianchi
models with $k^2<2$ (including standard matter satisfying standard
energy conditions) was completed by Kitada and Maeda
\cite{Kitada1993}. 
The governing differential equations in spatially homogeneous Bianchi
cosmologies containing a scalar field with an exponential potential
reduce to a dynamical system when appropriate
expansion- normalized variables are defined. This dynamical system was studied in
detail in \cite{Coley1997a} (where matter terms were not considered).

One particular solution that is of great interest is the flat, isotropic
power-law inflationary solution which occurs for $k^2<2$. This
power-law inflationary solution is known to be an attractor for all initially
expanding Bianchi models (except a subclass of the Bianchi type IX
models which will recollapse) \cite{Kitada1993,Coley1997a}. Therefore, all
of these models inflate forever; there is no exit from inflation and
no conditions for conventional reheating.

Recently cosmological models which contain both  
a scalar field with an exponential potential and a
barotropic perfect fluid
with a linear equation of state given by (1.2),
where $\gamma$ is in the physically relevant range
$2/3<\gamma \le 2$,
have come under heavy analysis.  One class of exact solutions found for
these models has the property that the energy density due to the
scalar field is proportional to the energy density of the perfect
fluid, and hence these models have been labelled matter scaling
cosmologies \cite{Wetterich1988a,Wands1993a,Ferreira1998a}.
These matter scaling solutions 
are spatially flat isotropic models and are known to be
late-time attractors (i.e., stable) in the subclass of
flat isotropic models \cite{Wands1993a}
and are clearly of physical interest.
In addition to the matter scaling solutions, curvature
scaling solutions \cite{vandenHoogen1999b} 
and anisotropic scaling solutions \cite{Coley1998b}
are also possible. A comprehensive analysis of
spatially homogeneous models with a perfect fluid
and a scalar field with an exponential potential has recently been
undertaken \cite{Billyard1999f}.

Although the exponential potential models are interesting models for a variety
of reasons, they have some shortcomings as {\em inflationary} models
\cite{Olive1990a,Linde1987a}.
While Bianchi models generically asymptote towards
the power-law inflationary model in which the matter terms are driven to
zero for $k^2<2$,
there is no graceful exit from this inflationary phase.
Furthermore, the scalar field cannot oscillate 
and so reheating cannot occur by the conventional scenario.
In recent work \cite{BC2000} interaction terms were included,
through which the energy of the scalar field is transferred to the
matter fields. These terms were found to
affect the qualitative behaviour of these models and, in particular, lead to
interesting inflationary behaviour.

\subsection{Isotropisation}

In the famous paper by Collins and Hawking \cite{CollinsHawking73} it 
was proven that within the set of spatially homogeneous cosmological 
models (which satisfy reasonable energy conditions)  those which 
approach isotropy at infinite times is of measure zero; that is, in 
general anisotropic models do not isotropize as they evolve to the 
future.  Since we presently observe the universe to be
highly isotropic,  we therefore need an explanation of why our universe 
has
evolved the way it has.  This problem, known as the isotropy problem, 
can be easily 
solved with an idea popularized by Guth \cite{Guth1981a}.  If  the early 
universe experiences a period of inflation, then all anisotropies are 
essentially pushed out of our 
present observable light-cone and are therefore not presently observed.  
The {\em Cosmic No-Hair Conjecture} asserts that under appropriate conditions,
any universe model will undergo a period of inflation and will 
consequently isotropize.

A signifigant amount of work on the Cosmic No-hair Conjecture has already 
been done for spatially homogeneous (Bianchi) cosmologies 
\cite{Wald83,Jensen87,Heusler91,Kitada92,Ibanez95}.  
For instance, Wald 
\cite{Wald83} has proven a version of the Cosmic No-Hair Conjecture for 
spatially homogeneous spacetimes with a positive 
cosmological constant; namely, he has shown   that all initially expanding 
Bianchi models asymptotically approach a spatially homogenous and 
isotropic model, except the subclass of  Bianchi type IX models which 
recollapse.  

Anisotropic models with scalar 
fields and with particular forms for the scalar field potential have also 
been investigated.  Heusler 
\cite{Heusler91} has analyzed the case in which the potential function 
passes 
through the origin and is concave up and,  like Collins and Hawking 
\cite{CollinsHawking73} has found that the only 
models that can possibly isotropize to the future are those of Bianchi 
types I, V, VII and IX. 
 
As noted above, 
Kitada and Maeda \cite{Kitada92,Kitada1993} have proven that if  $k<\sqrt{2}$, then all 
initially expanding 
Bianchi models except possibly those of type IX must isotropize.  Let us consider 
what happens in the case  $k> \sqrt{2}$.
In Ib\'a\~nez et al. \cite{Ibanez95} it was proven, using results from Heusler's paper 
\cite{Heusler91}, that the only models that can possibly isotropize when
$k>\sqrt{2}$
are those of Bianchi types I, V, VII, or IX.   Since the Bianchi I, V 
and VII$_0$ models are restricted classes of models, the only general spatially homogeneous 
models that can possibly isotropize are consequently of types VII$_h$ or IX.
Here we shall study the possible isotropization of 
 the Bianchi type VII$_h$ models when $k>\sqrt{2}$.

\subsubsection{The Bianchi VII$_h$ Equations}

 The Bianchi type VII$_h$ models belong to the Bianchi type B models 
 as classified by Ellis and MacCallum 
 \cite{EM}.  Hewitt and Wainwright 
 \cite{HW93} have derived the equations 
describing the evolution of the general Bianchi type B models.  
We shall utilize these equations, adjusted so that they 
describe a model with a minimally coupled scalar field $\phi$ with an 
exponential potential $V(\phi)=\Lambda e^{k\phi}$.   The energy-momentum 
tensor describing a minimally coupled scalar field is given by 
$$T_{ab}=\phi_{;a}\phi_{;b}-g_{ab}
\left(\frac{1}{2}\phi_{;c}\phi^{;c}+V(\phi)\right),$$
where, for a homogeneous scalar field, $\phi=\phi(t)$.  
In this case we can formally treat the  energy-momentum 
tensor as a perfect fluid with velocity vector 
$u^a={\phi^{;a}}/{\sqrt{-\phi_{;b}\phi^{;b}}}$, 
where the energy density, $\mu_{\phi}$, and the 
pressure, $p_{\phi}$, are given by 
\begin{eqnarray}
\mu_{\phi}&=&\frac{1}{2}\dot\phi^2+V(\phi),
\label{density}\\
p_\phi&=&\frac{1}{2}\dot\phi^2-V(\phi).
\end{eqnarray}
Our variables are the same as those used by Hewitt and Wainwright 
\cite{HW93}, with the addition of
\begin{equation}
x\equiv\sqrt{\frac{3}{2}}\frac{\dot\phi}{\theta},\qquad\qquad 
y\equiv\frac{\sqrt{3V}}{\theta}\label{dv1}
\end{equation}
to describe  the scalar field.   
We note that
$\Omega_\phi\equiv3\mu_\phi/\theta^2=x^2+y^2$.

The dimensionless evolution equations are then \cite{HW93}:
\begin{eqnarray}
\Sigma_+'         &=&  (q-2)\Sigma_+-2\tilde N, \label{sys1}\\
\tilde\Sigma'     &=& 2(q-2)\tilde\Sigma -4\Delta N_+-4\Sigma_+\tilde A, \\
\Delta'            &=&  2(q+\Sigma_+-1)\Delta+2(\tilde\Sigma-\tilde N)N_+,\\
\tilde A'          &=&  2(q+2\Sigma_+)\tilde A,\\
N_+'                 &=&  (q+2\Sigma_+)N_++6\Delta, \\
x'                &=&  (q-2)x - \sqrt{\frac{3}{2}}ky^2, \label{KG1}
\end{eqnarray}
where the prime denotes differentiation with respect to the new dimensionless time  $\tau$, where $\frac{dt}{d\tau}=\frac{3}{\theta}$, and 
\begin{eqnarray}
\tilde N       &=& \frac{1}{3}(N_+^2-\ell\tilde A), \\
q                 &=& 2\Sigma_+^2+2\tilde\Sigma+2x^2-y^2.
\end{eqnarray}
There also exists the constraint,
\begin{equation}
\tilde\Sigma\tilde N-\Delta^2-\tilde A\Sigma_+^2  = 0,
\end{equation}
and the equations are subject to the conditions
\begin{equation}
\tilde A \geq 0, \qquad\qquad \tilde\Sigma \geq 0, \qquad\qquad \tilde N  
\geq 0.
\end{equation}
The generalized Friedmann equation, written in dimensionless variables, becomes
\begin{equation}
 y^2=1-\Sigma_+^2-\tilde \Sigma -\tilde A-\tilde N-x^2,
\end{equation}
which serves to define $y$, and the evolution of $y$ is governed by
\begin{equation}
y'        =    (q+1+ \sqrt{\frac{3}{2}}kx)y.\label{KG2}
\end{equation}
Eqs. (\ref{KG1}) and (\ref{KG2}) are equivalent to the Klein-Gordon equation 
$\ddot\phi+\theta\dot\phi+V'(\phi)=0$ written in dimensionless variables.
The parameter $\ell=\frac{1}{h}$ defines the group parameter $h$ in the 
Bianchi  VII$_h$ models.

The variables $\Sigma_+$ and $\tilde \Sigma$ describe the shear 
anisotropy.  The variables $\tilde A$, $N_+$ and $\tilde N$ describe the 
spatial curvature of the models.  The variable $\Delta$ describes the 
relative orientation of the shear and spatial curvature eigenframes.

We are not interested here in the complete qualitative behaviour 
of the cosmological models \cite{ColeyBillyard} but simply whether the 
Bianchi VII$_h$ models isotropize to the future when $k>\sqrt{2}$.    
This question can be easily answered by examining the stability of the 
isotropic singular points of the above six-dimensional dynamical 
system (\ref{sys1}--\ref{KG1}).

\subsubsection{Stability Analysis}

All of the isotropic singular points lie in the invariant set {\it FRW} 
defined by $\{\Sigma_+=0,\tilde\Sigma=0,\tilde N=0, \Delta=0\}$.  Therefore, 
we shall find all of the isotropic singular 
points and  determine whether any are stable attractors or sinks \cite{vdH}.  

The singular point 
\begin{equation}
\Sigma_+=0,\tilde\Sigma=0,\Delta=0,A=1,N_+=\sqrt{\ell},x=0
\end{equation}
implies $y=0$ and
 represents the negatively curved Milne vacuum model.
The linearization of the dynamical system in the neighborhood of this singular point has 
eigenvalues
\begin{equation}
0,2,-2,-4,-2+4\sqrt{-\ell},-2+4\sqrt{-\ell}.
\end{equation}
Therefore, this singular point is a saddle.    

The singular point(s) 
\begin{equation}
\Sigma_+=0,\tilde\Sigma=0,\Delta=0,A=0,N_+=0,x=1 {\rm \ or \ } -1
\end{equation}
imply that $y=0$ and represent flat non-inflationary FRW model(s).
The eigenvalues in both cases are
\begin{equation}
0,0,2,2,4,6+\sqrt6 k.
\end{equation}
These singular points are unstable with an unstable manifold of at least dimension 4.

The singular point 
\begin{equation}
\Sigma_+=0,\tilde\Sigma=0,\Delta=0,A=0,N_+=0,x=-\frac{\sqrt{6}}{6}k \label{2.20}
\end{equation}
implies that $y=\sqrt{1-\frac{k^2}{6}}$. 
The eigenvalues   are
\begin{equation}
\frac{k^2-6}{2},\frac{k^2-6}{2},k^2-6,k^2-4,k^2-2,\frac{k^2-2}{2}
\end{equation}
For $k<\sqrt{2}$, this singular point represents the usual power-law inflationary 
attractor.  If $\sqrt{2}<k<2$, then the singular point has an unstable manifold of 
dimension 4.  If $2<k<\sqrt{6}$, then the singular point has an unstable manifold of 
dimension 3.  This singular point does not exist if $k>\sqrt{6}$.

The singular point 
\begin{equation}
\Sigma_+=0,\tilde\Sigma=0,\Delta=0,A=1-\frac{2}{k^2},N_+=\sqrt{\ell 
\left(1-\frac{2}{k^2}\right)},x=-\frac{\sqrt{6}}{3k}
\label{2.22}
\end{equation}
denoted $F$, implies that $y=\frac{2\sqrt{3}}{3k}$ and represents a 
non-inflationary negatively curved FRW model. 
The eigenvalues   are
\begin{eqnarray} 
-1+\sqrt{\frac{8}{k^2}-3},& &-1-\sqrt{\frac{8}{k^2}-3},\nonumber\\
-2+\frac{\sqrt{2}}{k}\sqrt{(k^2-4\ell k^2+8\ell)+E},&\qquad&
-2+\frac{\sqrt{2}}{k}\sqrt{(k^2-4\ell k^2+8\ell)-E},\\
-2-\frac{\sqrt{2}}{k}\sqrt{(k^2-4\ell k^2+8\ell)+E},& &
-2-\frac{\sqrt{2}}{k}\sqrt{(k^2-4\ell k^2+8\ell)-E},\nonumber
\end{eqnarray} 
where
$$E\equiv\sqrt{(k^2+4\ell k^2-8\ell)^2 + 32\ell(2-k^2)}.$$
After some   algebra it can be shown that if $k>\sqrt{2}$ (note that  $\ell>0$ in the 
Bianchi VII$_h$ models) then all of the eigenvalues have negative real parts.  Therefore, 
if $k>\sqrt{2}$, then this singular point is a stable attractor.  (Note that this 
singular point does not exist if $k<\sqrt{2}$.) In other words, there exists an open set 
of initial conditions in the set of anisotropic Bianchi VII$_h$ (with a scalar field and 
exponential potential) initial data for which the corresponding cosmological models 
asymptotically approach an isotropic and negatively curved FRW model.

\subsubsection{Discussion}

We have shown that within the set of all spatially homogeneous initial 
data, there exists an open set of initial data describing the  Bianchi type 
VII$_h$ models (having a scalar field with an exponential potential and 
$k>\sqrt{2}$) such that the models approach isotropy at infinite times.  
This compliments the results of Kitada and Maeda \cite{Kitada92,Kitada1993}, 
who showed that all ever-expanding spatially homogeneous models (including the 
Bianchi VII$_h$ models) with $k<\sqrt{2}$ approach isotropy to the future. 
In other words, there exists a set of spatially homogeneous initial data 
of non-zero measure for which models will isotropize to the future for 
all positive values of $k$.  Of course, there also exists a set of 
spatially homogeneous initial data of non-zero measure for which models 
will not isotropize to the future when $k>\sqrt{2}$ (e.g., the Bianchi VIII models).

If $k<\sqrt{2}$, then all models will inflate as they approach the   
power-law inflationary attractor represented by Eq. (\ref{2.20}).  For $k>\sqrt{2}$, the stable singular 
point $F$, given by Eq. (\ref{2.22}), which does not exist for 
$k<\sqrt{2}$, is isotropic and resides on the surface $q=0$.   
This means that the corresponding exact solution is  marginally 
non-inflationary.  However, this does not mean that the corresponding 
cosmological models are not inflating as they asymptotically approach 
this singular state.
As orbits approach $F$ they may have $q<0$ or $q>0$ (or even $q=0$) 
and consequently the models may or may not be inflating.  If they are 
inflating, then the rate of inflation is decreasing as $F$ is approached 
(i.e., $q\to 0$).  When $\sqrt{2}<k<\sqrt{8/3}$, we find that $F$ is 
node-like, hence there is an open set of models that inflate as they 
approach $F$ and an open set which do not.  When $k>\sqrt{8/3}$, $F$ 
is found to be spiral-like, and so it is expected that orbits 
experience regions of both $q<0$ and $q>0$ as they wind their way 
towards $F$.
As in Kitada and Maeda \cite{Kitada92,Kitada1993}, the inclusion of matter 
in the form of a perfect fluid is not expected to change the results 
of the analysis provided the matter satisfies  appropriate energy conditions.

\subsection {Stability of Matter Scaling Solutions}

Spatially homogeneous scalar field cosmological models
with an exponential potential and with barotropic matter
may also be important even if the 
exponential potential is too steep to drive inflation.  For example,
there exist `scaling solutions' in which the scalar field energy
density tracks that of the perfect fluid (so that at late
times neither field is negligible) \cite{Wetterich1988a}.
In particular, in \cite{Wands1993a}
a phase-plane analysis of the spatially flat 
FRW models showed that these scaling 
solutions are the unique late-time attractors whenever they 
exist.  The cosmological consequences of these scaling
models have been further studied in \cite{Ferreira1998a}.  For example, in 
such models a significant fraction of the current energy
density of the Universe may be contained in the homogeneous scalar field
whose dynamical effects mimic cold dark
matter;  the tightest constraint on these cosmological models comes from 
primordial nucleosynthesis bounds on any such relic density 
\cite{Wetterich1988a,Wands1993a,Ferreira1998a}.

Clearly these matter scaling models are of potential
cosmological significance. It is consequently of prime importance
to determine the genericity of such models by studying their
stability in the context of more general spatially homogeneous 
models.

\subsubsection{The Matter Scaling Solution} 

The governing equations for a scalar field with an 
exponential potential
$V = V_0 e^{k\phi}$  
 evolving in a flat
FRW model containing a separately conserved perfect fluid which satisfies
the barotropic equation of state 
$$p_\gamma = (\gamma -1) \mu_\gamma,$$
where  $0<\gamma < 2$ here, are given
by
\begin{eqnarray}
\label{m2} 
\dot{\theta} &= &-\frac{3}{2} (\gamma \mu_\gamma +  \dot{\phi}^2),  \\
\label{m3} 
\dot{\mu}_\gamma &= & - \gamma \theta \mu_\gamma,  \\
\label{m4} 
\ddot{\phi}&=& -\theta\dot{\phi}   - k V,  
\end{eqnarray}
subject to the Friedmann constraint
\be
\label{m5} 
\theta^2 = 3 (\mu_\gamma + \frac{1}{2} \dot{\phi}^2 + V),   
\ee
where an overdot denotes ordinary
differentiation with respect to time $t$, and units have been chosen
so that $8 \pi G =1$.  We note that the
total energy density of the scalar field is given by
Eq. (\ref{density}).

Defining $x$ and $y$ by Eq. (\ref{dv1})
and again using the logarithmic time variable, $\tau$, 
Eqs. (\ref{m2}) -- (\ref{m4}) can be written as the plane-autonomous system \cite{Wands1993a}:
\begin{eqnarray}
\label{m9} 
x^\p & = &-3x - \sqrt{\frac{3}{2}}  k y^2 + \frac{3}{2} x [2x^2 + 
\gamma (1-x^2 -y^2)],  \\
\label{m10} 
y^\p &=& \frac{3}{2}y \left[-\sqrt{\frac{2}{3}} -k x + 2x^2 + 
\gamma (1-x^2 -y^2)\right],  
\end{eqnarray}
where
 \be
 \label{m11} 
 \Omega \equiv \frac{3\mu_\gamma}{\theta^2}, \q \Omega_\phi \equiv 
\frac{3\mu_\phi}{\theta^2} = x^2 + y^2; \q  \Omega + \Omega_\phi =1,  
\ee
which implies that $0 \leq x^2 +y^2 \leq 1$ for $\Omega \geq 0$, so that the
phase-space is bounded.

A qualitative analysis of this plane-autonomous system
was given in \cite{Wands1993a}.  The well-known power-law inflationary
solution for $k^2<2$ \cite{Olive1990a, Kitada1993} corresponds to the singular point
$x = -k /\sqrt{6}$, $y = (1 - k^2/6)^{1/2}$ ($\Omega_\phi =1$, 
$\Omega =0$) of the system (\ref{m9})/(\ref{m10}), which is shown to be
stable (i.e., attracting) for $k^2 < 3 \gamma$ in the presence of a
barotropic fluid.  Previous analysis had shown that when $k^2<2$
this power-law inflationary solution is a global attractor in 
spatially homogeneous models in the absence of a perfect 
fluid (except for a subclass of Bianchi type IX models 
which recollapse).

In addition, for $\gamma > 0$ there exists a scaling solution
corresponding to the singular point
\be
\label{m13} 
x = x_0 = - \sqrt{\frac{3}{2}} \frac{\gamma}{k}, \q y = y_0 = [3
(2 - \gamma) \gamma/2 k^2]^{\frac{1}{2}},  
\ee 
whenever
$k^2 > 3\gamma$.  The linearization of system (\ref{m9})/(\ref{m10})
about the singular point (\ref{m13}) yields the two 
eigenvalues with negative real parts
\be
\label{m14} 
-\frac{3}{4}\left( 2-\gamma\right) \pm \frac{3}{4k} 
\sqrt{(2-\gamma)
[24\gamma^2-k^2(9\gamma-2)]} 
\ee
when $\gamma<2$.  The singular point is consequently stable (a
spiral for $k^2 > 24 \gamma^2/(9\gamma-2)$, else a node) so that
the corresponding cosmological solution is a late-time attractor in
the class of flat FRW models in which neither the scalar field nor the
perfect fluid dominates the evolution.  The effective equation of
state for the scalar field is given by
$$\gamma_\phi \equiv \frac{(\mu_\phi + p_\phi)}{ \mu_\phi} = 
\frac{2x^2_0}{x^2_0 + y^2_0} = \gamma, $$
which is the same as the equation of state parameter for the
perfect fluid.  The solution is referred to as a matter scaling solution since the energy
density of the scalar field remains proportional to that of 
the barotropic perfect fluid according to
$\Omega/\Omega_\phi = k^2/3\gamma -1$ \cite{Wetterich1988a}.  Since the
scaling solution corresponds to a singular point of the system
(\ref{m9})/(\ref{m10}) we note that it is a self-similar cosmological model \cite{cc}.

\subsubsection{Stability of the Matter Scaling Solution}

Let us study the stability of the matter scaling solution 
with respect to anisotropic and curvature perturbations within
the class of spatially homogeneous models \cite{BCH1999, Billyard1999f}.

{\em  Bianchi I models} 

In order to study the stability of the 
scaling solution with respect to shear perturbations we shall first 
investigate the class of anisotropic Bianchi I models, which 
are the simplest spatially homogeneous generalizations
of the flat FRW models and  have non-zero 
shear but zero three-curvature.  The governing equations in the Bianchi I 
models are 
Eqs. (\ref{m3}) and (\ref{m4}), and Eq. (\ref{m5}) becomes
\be
\label{m15} 
\theta^2 = 3 \left(\mu_\gamma + \frac{1}{2} \dot{\phi}^2 + V\right) 
+ \Sigma^2,
 \ee
where $\Sigma^2 \equiv 3\Sigma^2_0 R^{-6}$ is the contribution due 
to the shear, where $\Sigma_0$ is a constant and $R$ is the  
scale factor.  Eq. (\ref{m2}) is replaced by the time derivative of 
Eq. (\ref{m15}).

Using the definitions (\ref{dv1}) and (\ref{m11}) we can
deduce the governing ODEs. Due to 
the $\Sigma^2$ term in (\ref{m15}) we can no longer use this equation
to substitute for $\mu_\gamma$ in the remaining equations, and 
we consequently obtain the three-dimensional autonomous 
system:
\begin{eqnarray}
\label{m16} 
x^\p &=& -3x - \sqrt{\frac{3}{2}} k y^2 + \frac{3}{2} x [2 + (\gamma -2) 
\Omega -2y^2],  \\
\label{m17} 
y^\p &=& \frac{3}{2}y \left\{ \sqrt{\frac{2}{3}} k x +2 + (\gamma -2) 
\Omega
-2y^2\right\},  \\
\label{m18} 
\Omega^\p &=& 3 \Omega \{(\gamma -2)(\Omega -1) -2y^2\},  
\end{eqnarray}
where Eq. (\ref{m15}) yields
 \be
 \label{m19} 
 1 - \Omega -x^2 -y^2 = \Sigma^2 \theta^{-2} \geq 0,  
\ee
so that we again have a bounded phase-space.

The matter scaling solution, corresponding to the flat FRW 
solution, is now represented by the singular point
\be
\label{m20} 
x = x_0, ~~y = y_0, ~~\Omega =1 - \frac{3 \gamma}{k^2}.  
\ee
The linearization of system (\ref{m16}) -- (\ref{m18}) about the singular 
point (\ref{m20}) yields three eigenvalues, two of which are given by (\ref{m14})
and the third has the value $-3(2-\gamma)$, all with negative real
parts when $\gamma<2$.  Consequently the scaling solution is stable to 
Bianchi type I shear perturbations.

{\em  Curved FRW models}  

In order to study the stability of the scaling solution
with respect to curvature perturbations we shall first consider the
class of FRW models, which have curvature but no shear. Again
Eqs. (\ref{m3}) and (\ref{m4}) are valid, but in this case Eq. (\ref{m5}) becomes
\be
\label{m21} 
\theta^2 = 3(\mu_\gamma + \frac{1}{2} \dot{\phi}^2 +V) + K, 
\ee
where $K \equiv -9 k R^{-2}$ and $k$ is a constant that can be scaled to $0$, $\pm 1$.
Eq. (\ref{m2}) is again replaced by the time derivative of Eq.
(\ref{m21}).

As in the previous case we cannot use Eq. (\ref{m21}) to replace $\mu_\gamma$, and 
using the definitions
(\ref{dv1}) and (\ref{m11}) we 
obtain the three-dimensional autonomous system:
\begin{eqnarray}
\label{m22} 
x^\p &=&-3x - \sqrt{\frac{3}{2}}k y^2 + \frac{3}{2}x 
\left[ \left( \gamma - \frac{2}{3} \right) \Omega + \frac{2}{3} 
(1 + 2x^2 -y^2)  \right],  \\
\label{m23} 
y^\p &=& \frac{3}{2} y \left\{\sqrt{\frac{2}{3}} k x + 
\left( \gamma -\frac{2}{3}  \right) \Omega + \frac{2}{3} (1+2x^2 -y^2) 
\right\},  \\
\label{m24} 
\Omega^\p &=& 3 \Omega \left\{\left(\gamma - \frac{2}{3}\right)
(\Omega -1) + \frac{2}{3} (2x^2 -y^2)  \right\},  
\end{eqnarray}
where
\be
\label{m25} 
1 - \Omega - x^2 -y^2 = KH^{-2}.  
\ee
The phase-space is bounded for $K =0$ or $K<0$, but not for $K > 0.$

The matter scaling solution again corresponds to the singular point (\ref{m20}).
The linearization of system (\ref{m22}) -- (\ref{m24}) about this singular point
yields the two eigenvalues with negative real parts given by (\ref{m14}) and
the eigenvalue $(3 \gamma -2)$. Hence the scaling solution is only
stable for $\gamma < \frac{2}{3}$.  For $\gamma > \frac{2}{3}$ the
singular point (\ref{m20}) is a saddle with a two-dimensional stable
manifold and a one-dimensional unstable manifold.

Consequently the scaling solution is unstable
to curvature perturbations in the case of realistic
matter $(\gamma \geq 1)$; i.e., the scaling solution is no longer
a late-time attractor in this case.  However, the scaling solution
does correspond to a  singular point of the governing 
autonomous system of ODEs
and hence there are cosmological models that can spend
an arbitrarily long time `close' to this solution.  Moreover,
since the curvature of the Universe is presently constrained
to be small by cosmological observations, it is possible
that the scaling solution could be important in 
the description of our actual Universe.  That is, not enough
time has yet elapsed for the curvature instability to
have effected an appreciable deviation from the flat
FRW model (as in the case of the standard perfect fluid
FRW model). Hence the scaling solution may still be of physical 
interest.

\noindent
{\em Bianchi VII$_h$ models} 

To
further study the  significance of the scaling solution it is important to determine its
stability within a general class of spatially homogeneous models such as the
(general) class of Bianchi type VII$_h$ models (which are perhaps the most
physically relevant models since they can be regarded as
generalizations of the  negative-curvature  FRW models). 
The Bianchi VII$_h$ models are sufficiently complicated 
that a simple coordinate approach
(similar to that given above) is not desirable.
In subsection 3.2.1 the  Bianchi VII$_h$ spatially homogeneous 
models with a minimally coupled scalar field 
with an exponential
potential (but without a barotropic perfect fluid) were studied by
employing a group-invariant orthonormal frame approach
with expansion-normalized state variables governed by
a set of dimensionless evolution equations (constituting
a `reduced' dynamical system) with respect to a 
dimensionless time subject to a non-linear constraint \cite{HW93}.
A barotropic perfect fluid can easily be included \cite{Billyard1999f}.

The reduced dynamical system is seven-dimensional (subject to a
constraint).  The scaling solution is again a  singular point
of this seven-dimensional system. This singular point, which only
exists for $k^2>3\gamma$, has two eigenvalues given by (\ref{m14}) which
have negative real parts for $\gamma<2$, two eigenvalues
(corresponding to the shear modes) proportional to $(\gamma -2)$ which
are also negative for $\gamma <2$, and two eigenvalues (essentially
corresponding to curvature modes) proportional to $(3 \gamma -2)$
which are negative for $\gamma <\frac{2}{3}$ and positive for $\gamma
>\frac{2}{3}$ \cite{BCH1999}.  The remaining eigenvalue (which also corresponds
to a curvature mode) is equal to $3\gamma-4$.  Hence for $\gamma <
\frac{2}{3}$ ($k^2 > 3 \gamma$) the scaling solution is again
stable.  However, for realistic matter ($\gamma \geq 1$) the
corresponding singular point is a saddle with a 
four- or five-dimensional stable manifold (depending upon whether 
$\gamma>4/3$ or $\gamma<4/3$, respectively).

\section{String Models}
\setcounter{equation}{0}

There has been considerable interest recently 
in the cosmological implications of string theory. String theory 
introduces significant modifications to
the standard, hot big bang model based on conventional 
Einstein gravity.  
Early-universe cosmology provides one of the few environments 
where the predictions of the theory can be 
quantitatively investigated.
The evolution of the very early universe 
below the string scale is determined by ten--dimensional 
supergravity theories  \cite{Green1987a,eff}. 
All theories of this type contain a dilaton, a graviton 
and a two--form potential 
in the Neveu--Schwarz/Neveu--Schwarz (NS--NS) bosonic sector. 
If one considers a Kaluza--Klein 
compactification from ten dimensions onto an 
isotropic six--torus of radius $e^{\beta}$, 
the effective action is given by 
\begin{equation}
\label{NSaction}
S=\int d^4 x \sqrt{-g} e^{-\Phi} \left[ R 
+\left( \nabla \Phi \right)^2 -6 \left( \nabla \beta \right)^2
-\frac{1}{12} H_{\mu\nu\lambda} H^{\mu\nu\lambda} 
\right]  ,
\end{equation}
where the moduli fields arising from the 
compactification of the form--fields on the internal 
dimensions and 
the graviphotons originating from the 
compactification  of the metric have been neglected \cite{lower}. 
In Eq. (\ref{NSaction}),  
$R$ is the Ricci curvature of the spacetime with metric $g_{\mu\nu}$ and 
$g\equiv {\rm det}g_{\mu\nu}$, the dilaton field, $\Phi$, 
parametrizes the string coupling, $g_s^2 \equiv  
e^{\Phi}$, and $H_{\mu\nu\lambda} \equiv 
\partial_{[\mu} B_{\nu\lambda ]}$ is the field strength 
of the  two--form potential $B_{\mu\nu}$. The volume of the internal 
dimensions is parametrized by the modulus field, $\beta$.

In four dimensions, the  three--form 
field strength is dual to a one--form:  
\begin{equation}
\label{sigma}
H^{\mu\nu\lambda} \equiv e^{\Phi} \epsilon^{\mu\nu\lambda\kappa}
\nabla_{\kappa} \sigma   ,
\end{equation}
where $\epsilon^{\mu\nu\lambda\kappa}$ is the covariantly 
constant four--form. In this dual formulation, the 
field equations (FE) can be derived from the action 
\begin{equation}
\label{sigmaaction}
S=\int d^4 x \sqrt{-g} e^{-\Phi} \left[ 
R +\left( \nabla \Phi  \right)^2  -6 \left( \nabla \beta \right)^2 
-\frac{1}{2} 
e^{2 \Phi} \left( \nabla \sigma \right)^2 \right]   ,
\end{equation}
where $\sigma$ is interpreted as a pseudo--scalar `axion' field
\cite{sen}. 

It can be shown 
that the action (\ref{sigmaaction}) 
is invariant under a global ${\rm SL}(2, R)$ transformation on the 
dilaton and axion fields \cite{sen}. 
The general  
FRW cosmologies derived from Eq. 
({\ref{sigmaaction}) have been found by employing this symmetry \cite{clw}. 
However, the symmetry is broken 
when a cosmological constant is present \cite{kms} and the general 
solution is not known in this case. The purpose here is to determine  
the general structure of the phase space of solutions for the wide class 
of string cosmologies that contain a cosmological constant in 
the effective action. 
This is particularly 
relevant in light of recent high redshift observations 
that indicate a vacuum energy density may be 
dominating the large--scale dynamics of the universe at the present epoch 
\cite{line}. 

A cosmological constant 
may arise in a number of different contexts and we consider a 
general action of the form 
\begin{equation}
\label{generalaction}
S=\int d^4 x \sqrt{-g} \left\{ e^{-\Phi} \left[ R 
+\left( \nabla \Phi \right)^2 -6 \left( \nabla \beta \right)^2
-\frac{1}{2} e^{2\Phi} \left( \nabla \sigma \right)^2 
-2\Lambda \right] - \Lambda_{\rm R} \right\}   .
\end{equation}
The constant,  
$\Lambda$, is determined by the central charge deficit 
of the string theory and may be viewed as a cosmological constant 
in the gravitational sector of the theory.
In principle, it may take arbitrary values  
if the string is coupled to an appropriate  conformal 
field theory. Such a term may also have an origin in terms of 
the reduction of higher degree form--fields \cite{kaloper1}. 
The constant,  $\Lambda_{\rm R}$, represents 
a phenomenological cosmological constant in the matter sector. 
Since it  
does not couple directly to the dilaton field, it may be viewed in a stringy 
context as a Ramond--Ramond (RR) degree of freedom 
(a 0--form) \cite{lukas}. 
Such a cosmological constant may also be interpreted 
as the potential energy of a scalar field 
that is held in a false vacuum state. 

We shall include the 
combined effects of the axion, modulus and dilaton fields, thereby  
extending previous qualitative analyses where 
one or more of these terms was neglected 
\cite{gp,kmol,emw,kmo}.   A
full stability analysis can be performed for all models 
by rewriting the FE 
in terms of a set of compactified variables.  
As usual, units in which $c=8\pi G=1$ will be utilized throughout. 

\subsection{Cosmological Field Equations}

When $\Lambda_{\rm R} =0$, the spatially flat FRW cosmological 
FE derived from action (\ref{generalaction}) are given by 
\begin{eqnarray}
\label{ns1}
2\ddot{\alpha} -2\dot{\alpha}\dot{\varphi} -\dot{\sigma}^2 e^{2\varphi +6
\alpha}  &=&0\\
\label{ns2}
2\ddot{\varphi} -\dot{\varphi}^2 -3\dot{\alpha}^2 - 6 \dot{\beta}^2 
+\frac{1}{2} 
\dot{\sigma}^2 e^{2\varphi +6 \alpha}  + 2\Lambda &=&0 \\
\label{ns3}
\ddot{\beta} -\dot{\beta} \dot{\varphi} &=&0 \\
\label{ns4}
\ddot{\sigma} +\dot{\sigma} \left( \dot{\varphi} +6\dot{\alpha} \right) &=&0 ,
\end{eqnarray}
where $\varphi \equiv \Phi -3\alpha$ defines the `shifted' 
dilaton field,  $R(t) \equiv e^{\alpha}$ is the scale 
factor of the universe 
and a dot denotes differentiation with respect to 
cosmic time, $t$. The generalized Friedmann constraint equation is 
\begin{equation}
\label{nsfriedmann}
3\dot{\alpha}^2 -\dot{\varphi}^2 +6 \dot{\beta}^2 +\frac{1}{2} \dot{\sigma}^2 
e^{2\varphi + 6 \alpha} +2\Lambda =0   .
\end{equation}

A number of exact solutions to Eqs. (\ref{ns1})--(\ref{nsfriedmann}) 
are known when one or more of the degrees of freedom are trivial; 
these solutions lie in
the invariant sets of the full phase space. 
The `dilaton--vacuum' solutions, where only the dilaton field 
is dynamically important, are given in \cite{pbb};
there 
is a curvature singularity in these solutions at $t=0$. 
In the pre--big bang inflationary 
scenario, the pre--big bang phase corresponds to 
the range $t<0$ and the post--big 
bang phase to the solution for $t>0$. 
The `dilaton--moduli--vacuum' solutions have 
$\dot{\sigma} = \Lambda =0$.  
The general solution with $\Lambda =0$
is the `dilaton--moduli--axion' solution \cite{clw}; 
this cosmology asymptotically approaches a
dilaton--moduli--vacuum solution in the limits 
of high and low spacetime curvature. 
The axion field induces a smooth transition between these two 
power-law solutions. 
The solutions where only the axion 
field is trivial and $\Lambda >0$ are specific 
cases of the `rolling radii' solutions found by Mueller 
\cite{mu}.
The corresponding solutions for $\Lambda < 0$ are related  
by a redefinition.
Finally, there exists the `linear dilaton--vacuum' solution 
where $\Lambda >0$ \cite{myers}. This solution is static and the dilaton 
evolves linearly with time.

\subsection{Qualitative Analysis of the NS--NS Fields}

For an arbitrary central charge deficit,  
the FE (\ref{ns1})--(\ref{nsfriedmann}) may 
be written as an autonomous system of  ODEs: 
\begin{eqnarray}
\label{ns1a}
\dot{h} &=&\psi^2 + h \psi -3h^2 -N  -2\Lambda \\
\label{ns2a}
\dot{\psi} &=&3h^2 +N \\
\label{ns3a}
\dot{N} &=& 2N \psi \\
\label{ns4a}
\dot{\rho} &=& -6h \rho \\
\label{nsfriedmanna}
3h^2 &-&\psi^2 +N +\frac{1}{2} \rho +2\Lambda =0   ,
\end{eqnarray}
where we have defined the new variables
\begin{equation}
\label{beta}
N \equiv 6\dot{\beta}^2 , \qquad \rho \equiv \dot{\sigma}^2  e^{2\varphi 
+6\alpha} , 
\qquad \psi \equiv \dot{\varphi} , \qquad h \equiv \dot{\alpha}  ,
\end{equation}
and $h=\theta/3$, where $\theta$ is the expansion scalar defined earlier.
The variable $\rho$ may be interpreted as the effective 
energy density of the pseudo--scalar axion field \cite{kmo}.
It follows from Eq.  (\ref{ns2a}) that $\psi$ is a
monotonically increasing function of time 
and this implies that the singular
points of the system of ODEs must be located either at zero or infinite 
values of $\psi$.
In addition, due to the existence of a monotone function,
it follows that there are no periodic or recurrent orbits
in the corresponding phase space \cite{WE, Hale}. The sets
$\Lambda =0$ and $\rho =0$ are invariant sets. In particular, 
the exact solution for
$\Lambda =0$ divides the phase space and the
orbits do not cross from positive to negative $\Lambda$.

We  must consider the cases where $\Lambda < 0$ and 
$\Lambda >0$ separately. 
In the  case where the central charge deficit 
is negative, $\Lambda < 0$, 
it proves convenient 
to employ the generalized Friedmann constraint  equation 
(\ref{nsfriedmanna}) to eliminate the  
modulus field. 
We may  compactify the phase space by
normalizing with 
$\sqrt{\psi^2 -2\Lambda}$ and we define
a new time variable by
\begin{equation}
\label{tau}
\frac{d}{dT} \equiv \frac{1}{\sqrt{\psi^2 -2\Lambda}} \frac{d}{dt}  .
\end{equation}
The governing equations reduce to a three--dimensional system
of autonomous ODEs.
The singular points all lie on one of the two
lines of non-isolated singular points (or one-dimensional singular sets)
$L_\pm$.
On the line $L_+$ the
singular points are either saddles or
local sinks, and on the line
$L_-$ the singular points are local
sources or saddles. The dynamics
is very simple due to the existence of (two) monotonically increasing functions. 
A full analysis is given in \cite{BCL}.
Henceforward, let us consider the case $\Lambda >0$.

\subsubsection{Models with Positive Central Charge Deficit}}

In the case where the central charge deficit is 
positive, $\Lambda >0$, we choose the 
normalization
\begin{equation}
\label{epsilon}
\epsilon \equiv \left( 3h^2 +\frac{1}{2} \rho +N +2\Lambda \right)^{1/2}  .
\end{equation}
The generalized Friedmann constraint equation (\ref{nsfriedmanna}) now takes the simple form
\begin{equation}
\label{unity}
\frac{\psi^2}{\epsilon^2} =1
\end{equation}
and may be employed to eliminate $\psi$. 
Since by definition 
$\epsilon \ge 0$, specifying one of the roots $\psi/\epsilon
=\pm 1$ corresponds to choosing the sign of $\psi$. However, it follows 
from the definition in Eq. (\ref{beta}) that changing the sign of $\psi$ 
is related to a time reversal of the dynamics. 
In what follows, we shall consider the case $\psi/\epsilon=+1$; the case $\psi/\epsilon=-1$
is qualitatively similar.

Introducing the new normalized variables
\begin{equation}
\label{nsnewa}
\mu \equiv \frac{\sqrt{3}h}{\epsilon} , 
\qquad \nu \equiv \frac{\rho}{2\epsilon^2} , \qquad 
\lambda \equiv \frac{N}{\epsilon^2} 
\end{equation}
and a new dynamical variable
\begin{equation}
\frac{d}{dT} \equiv \frac{1}{\sqrt{3} \epsilon} \frac{d}{dt}
\end{equation}
transforms Eqs. (\ref{ns1a})--(\ref{ns4a}) to the three--dimensional 
autonomous system: 
\begin{eqnarray}
\label{s3}
\frac{d \mu}{dT} &=& \nu +\frac{\mu}{\sqrt{3}}  
\left[ 1-\mu^2 -\lambda \right] \label{dmu} \\
\label{s4}
\frac{d \nu}{dT} &=& -2 \nu \left[ \mu + 
\frac{1}{\sqrt{3}} 
\left( \lambda + \mu^2 \right) \right] \label{dnu}\\
\frac{d \lambda}{d T} &=& \frac{2}{\sqrt{3}} 
\lambda \left( 1 - \mu^2 -\lambda \right) \label{dlambda}  .
\end{eqnarray}
The phase space variables are bounded, $0 \le 
\{ \mu^2 , \nu , \lambda \} \le 1$, and  satisfy $\mu^2 +\nu
+\lambda \le 1$.  The sets $\nu=0$ and $\lambda=0$ are invariant sets
corresponding to $\rho=0$ and $\dot{\beta}=0$, respectively.  In addition,
$\mu^2+\nu+\lambda=1$ is an invariant set corresponding to
$\Lambda=0$.  We note that the right-hand side of Eq. (\ref{dlambda}) is
positive-definite and this simplifies the
dynamics considerably.

The singular points of the system (\ref{dmu})-(\ref{dlambda}) 
consist of the isolated singular point
\begin{equation}
C:\mu=\nu=\lambda=0,
\end{equation}
and the line of non-isolated singular points
\begin{equation}
V:\nu=0, \lambda=1-\mu^2 \ \ \mbox{($\mu$ arbitrary)}.
\end{equation}
The eigenvalues associated with $C$ are
$\lambda_1=1/\sqrt{3}$, $\lambda_2=2/\sqrt{3}$ and
$\lambda_3=0$. Since $C$ is an isolated singular point,  it is therefore
non-hyperbolic.  However, a simple analysis shows that
it is a global source.
The eigenvalues associated with $V$ are:
\begin{equation}
\lambda_1=-2\left( \mu+\frac{1}{\sqrt{3}} \right)  , 
 \quad \lambda_2=-\frac{2}{\sqrt{3}} 
\end{equation}
and the third eigenvalue is zero since $V$ is a one--dimensional set.
Therefore,  on $V$ the singular points are saddles for $\mu\in
[-1,-1/\sqrt{3})$ and local sinks for $\mu\in(-1/\sqrt{3},1]$.  

It is also 
instructive to consider the dynamics on the boundary corresponding to 
$\lambda=0$, since the case $N=0$ is of physical interest in its own right
as a four--dimensional model.  In this case 
the ODEs reduce to the two-dimensional system:
\begin{eqnarray}
\label{2d1}
\frac{d\mu}{dT} &=& \nu+\frac{\mu}{\sqrt{3}}\left(1-\mu^2\right) \\
\label{2d2}
\frac{d\nu}{dT} &=& -2\mu\nu \left[ 1 +\frac{1}{\sqrt{3}}\mu \right].
\end{eqnarray}
The singular
points and their corresponding eigenvalues are:
\begin{eqnarray}
C:& \mu=    \nu=0; & \lambda_1=\frac{1}{\sqrt{3}}, \lambda_2=0 \\
S:& \mu=-1, \nu=0; & \lambda_1=-\frac{2}{\sqrt{3}}, 
\lambda_2=2\left(1-\frac{1}{\sqrt{3}}                       \right) \\
A:& \mu= 1, \nu=0; & \lambda_1=-\frac{2}{\sqrt{3}}, 
\lambda_2=-2\left(1+\frac{1}{\sqrt{3}}                       \right).
\end{eqnarray}
Point $C$ is a non-hyperbolic singular point; however, by 
changing to polar 
coordinates we find that $C$ is a repeller with an invariant ray 
$\theta=\tan^{-1}(-\sqrt{3})$.  
The saddle $S$ and the attractor $A$ lie on the 
line $V$.\\

\subsubsection{Discussion}

Let us briefly summarize the dynamics in the case $\Lambda <0$, which was studied in \cite{BCL}.
When the modulus field is frozen  the universe contracts from
a singular initial state. The orbits are past--asymptotic to a dilaton--vacuum
solution, and so the axion is negligible and the kinetic energy
of the dilaton is dominant.  As the collapse
proceeds, however, the axion becomes dynamically more important and
eventually induces a bounce.  In the case of vanishing $\Lambda$,
these equations imply that the future attractor would correspond to
a dilaton--vacuum solution. However, the combined effect of
the axion and the central charge deficit is to cause the universe to evolve
towards a singularity with $\dot{\alpha} \rightarrow
+\infty$ in a finite time.  This behaviour is different from that found
when no axion field is present in which there is
no bounce \cite{kmol,mu}.
The inclusion of a modulus field leads to
a line of sources and a line of sinks. The 
axion field is dynamically negligible in the neighbourhood 
of the singular points. Moreover, a bouncing cosmology 
is no longer inevitable and there exist solutions that 
expand to infinity in a finite time. The solutions are asymptotic to
the dilaton--moduli--vacuum solutions  near the lines 
$L_{\pm}$. 

In the case of a positive $\Lambda$ discussed here,  the isolated singular point $C$ 
corresponds to the `linear dilaton--vacuum' solution  
\cite{kmol,myers}. When 
the modulus is frozen, all trajectories evolve away from $C$ 
towards the point $A$ and approach a superinflationary  
dilaton--vacuum solution
defined over $t<0$. Some of the orbits evolving away from $C$ 
represent contracting cosmologies and the 
effect of the axion is to reverse 
the collapse in all these cases. 
For the rolling modulus solutions, the orbits tend to 
dilaton--moduli--vacuum solutions as they approach the attractors 
(the sinks on $V$). 
As in the case of a negative 
central charge deficit, the critical 
value $\mu^2 =1/3$ corresponds to the case where 
$\dot{\alpha}^2=\dot{\beta}^2$,
representing  the isotropic, ten--dimensional cosmology 
$(\dot{\alpha} = \dot{\beta})$ 
and its dual solution $(\dot{\alpha} =-\dot{\beta})$. 
In the latter solution, the 
ten--dimensional dilaton field, $\hat{\Phi} \equiv \Phi +6\beta$, 
is constant. 
 The other boundary of $V$ is the point $A$ representing the case 
where the kinetic energy of the modulus field vanishes. 
We note that the qualitative behaviour of models with
$\psi<0$ is similar.

\subsection{Qualitative Analysis of the RR Sector}

When a cosmological constant is introduced into the 
matter sector of Eq. (\ref{generalaction}) and the central 
charge deficit vanishes (i.e., $\Lambda=0$), the FE are given by  
\begin{eqnarray}
\label{r1}
\ddot{\alpha}  &=&\dot{\alpha} \dot{\varphi} +\dot{\varphi}^2 -3\dot{\alpha}^2 
-6 \dot{\beta}^2 -\frac{3}{2} \Lambda_{\rm R} e^{\varphi +3\alpha} \\
\label{r2}
\ddot{\varphi} &=&3\dot{\alpha}^2  + 6 \dot{\beta}^2 +\frac{1}{2} 
\Lambda_{\rm R} 
e^{\varphi +3\alpha}  \\
\label{r3}
\ddot{\sigma} &=&-(\dot{\varphi}  +6 \dot{\alpha}  ) \dot{\sigma} \\
\label{r4}
\ddot{\beta} &=&\dot{\beta} \dot{\varphi} 
\end{eqnarray} 
and the generalized Friedmann constraint equation takes the form   
\begin{equation}
\label{frr}
3\dot{\alpha}^2 -\dot{\varphi}^2 +6 \dot{\beta}^2  +\frac{1}{2} \dot{\sigma}^2 
e^{2\varphi +6 \alpha} + 
\Lambda_{\rm R} e^{\varphi +3\alpha} =0   .
\end{equation} 
Eqs. (\ref{r1})--(\ref{frr}) may be simplified by introducing the new 
time coordinate
\begin{equation}
\frac{d}{d \tilde t} \equiv e^{-(\varphi +3\alpha )/2} \frac{d}{dt}
\end{equation}
and employing the generalized Friedmann constraint equation (\ref{frr}) 
to eliminate the axion field. The remaining  
FE are then given by 
\begin{eqnarray}
\label{r5}
\alpha'' &=& \varphi'^2 -\frac{9}{2} \alpha'^2 +\frac{1}{2} \alpha'
\varphi' 
-6 \beta'^2 -\frac{3}{2} \Lambda_{\rm R} \\
\label{r6}
\varphi'' &=&3\alpha'^2 +6\beta'^2 -\frac{1}{2} \varphi'^2 -
\frac{3}{2} \alpha' \varphi' +\frac{1}{2} \Lambda_{\rm R}  \\
\label{r7}
\beta'' &=& \frac{1}{2} \beta' \left( 
\varphi' -3 \alpha'  \right) ,
\end{eqnarray} 
where in this section a prime denotes differentiation with 
respect to $\tilde t$. 

\subsubsection{Positive Cosmological Constant}

When $\Lambda_{\rm R} > 0$, 
we express Eqs. (\ref{r5})--(\ref{r7}) as an autonomous system by 
defining  
\begin{equation}
\label{4.10}
h \equiv \alpha',  \qquad \psi \equiv \varphi' , \qquad 
N \equiv \beta'  .
\end{equation}
Eq. (\ref{frr}) then implies that  
\begin{equation}
\psi^2 \ge 3h^2 +6 N^2 +\Lambda_{\rm R} \ge 0
\end{equation} 
and consequently we may normalize using $\psi$. We therefore 
define
\begin{eqnarray}
\label{x}
x &\equiv& \frac{\sqrt{3} h}{\psi} \\
\label{y}
y &\equiv& \frac{6 N^2}{\psi^2} \\
\label{z}
z &\equiv& \frac{\Lambda_{\rm R}}{\psi^2} \\
\label{theta}
\frac{d}{d\Theta} &\equiv& \frac{1}{\psi} \frac{d}{d\tilde t},
\end{eqnarray}
and assume that $\psi>0$ ($\psi<0$ is again related to time-reversal).

The resulting three--dimensional autonomous system is:
\begin{eqnarray}
\frac{dx}{d\Theta} &= & (x+\sqrt{3})[1-x^2-y-z] +\frac{1}{2} z[x-\sqrt{3}] 
\label{xp} \\
\frac{dy}{d\Theta} &= & 2y\left\{[1-x^2-y-z]+\frac{1}{2}z\right\} 
\label{yp} \\
\frac{dz}{d\Theta} &= & 2z\left\{[1-x^2-y-z] 
-\frac{1}{2}(1-z-\sqrt{3}x)\right\} \label{zp}   .
\end{eqnarray}
It follows from the definitions (\ref{x})--(\ref{z}) 
that the phase space is
bounded with $0\leq \{ x^2,y,z\}\leq 1$, subject to 
the constraint $1-x^2-y-z\geq0$.  The
invariant set $1-x^2-y-z=0$ corresponds to a zero axion field.  The
dynamics of the system (\ref{xp})--(\ref{zp}) is determined primarily
by the dynamics in the invariant sets $y=0$ and $z=0$. 
These correspond
to a zero modulus field and a zero $\Lambda_{\rm R}$, respectively. 
The dynamics is also determined by
the fact that the right-hand side of Eq. (\ref{yp}) is positive--definite
so that $y$ is a monotonically increasing function. This 
guarantees that there are no closed or
recurrent orbits in the three-dimensional phase space. 

\subsubsection{Four--Dimensional Model}

In the invariant set $y=0$, where the modulus field is trivial, 
the system (\ref{xp})-(\ref{zp}) reduces to the following plane system:
\begin{eqnarray}
\label{dxdTh}
\frac{dx}{d\Theta} &= & (x+\sqrt{3})[1-x^2-z] +\frac{1}{2} z[x-\sqrt{3}] \\
\label{dzdTh}
\frac{dz}{d\Theta} &= & 2z\left\{[1-x^2-z] 
-\frac{1}{2}(1-z-\sqrt{3}x)\right\} .
\end{eqnarray}
The singular points and their associated eigenvalues are given by
\begin{eqnarray}
S_1: & x=-1, z=0; & \lambda_1=2(\sqrt{3}-1), \quad \lambda_2=-(1+\sqrt{3}) \\
S_2: & x= 1, z=0; & \lambda_1=-2(\sqrt{3}+1), \quad \lambda_2= (\sqrt{3}-1) \\
F: & x=-\frac{1}{3\sqrt{3}}, z=\frac{16}{27}; & \lambda_{1,2}= \frac{1}{3} \pm 
\frac{i}{9}\sqrt{231}.
\end{eqnarray}
The points $S_1$ and $S_2$ are saddles and $F$ is a repelling focus.

In the invariant set $1-x^2-z=0$, corresponding to the case of a
zero axion field, Eqs. (\ref{dxdTh}) and (\ref{dzdTh}) reduce to the
single ODE
\begin{equation}
\label{invariantset}
\frac{dx}{d\Theta} = \frac{1}{2}\left(1-x^2\right)\left(x-\sqrt 3\right),
\end{equation}
which can be integrated to yield an exact solution (in terms of $\Theta$ time).

\subsubsection{Ten--dimensional Model}

In the full system (\ref{xp})-(\ref{zp}) with a non--trivial modulus 
field, there exists the 
isolated singular point $F:x=-1/(3\sqrt{3}), y=0, z=16/27$
with the associated eigenvalues $\lambda_{1,2}$ and $\lambda_3=4/3$. 
This implies that 
$F$ is a global source. All of the remaining singular
points belong to the one--dimensional set
\begin{equation}
W: y=1-x^2, z=0 \ \ \mbox{($x$ arbitrary)}
\end{equation}
and the associated eigenvalues are given by
\begin{equation}
\lambda_1=-2\sqrt 3\left(x+\frac{1}{\sqrt{3}}\right), \quad 
\lambda_2=\sqrt 3\left(x-\frac{1}{\sqrt{3}} \right), 
   \quad (\lambda_3=0).
\end{equation}
$W$ lies in the invariant set $z=0$ on the boundary $y=1-x^2$.  Points
on $W$ with $x\in (-1/\sqrt{3},1/\sqrt{3})$ are local sinks, while the
remaining points are saddles in the full three-dimensional phase
space.  (In the invariant set $z=0$ 
singular points with $x\in
[-1,-1/\sqrt{3})$ are repelling and those with $x\in(-1/\sqrt{3},1]$
are attracting). 

We note that there exists an exact solution of Eqs. (\ref{xp})-(\ref{zp}) with non-trivial modulus field with
\begin{equation}
x=-\frac{1}{3\sqrt 3}= \mbox{constant}, \label{xfixed}
\end{equation}
and 
\begin{equation}
y=-\frac{13}{8}\left(z-\frac{16}{27}\right),\label{yline}
\end{equation}
whence
\begin{equation}
\frac{dz}{d\Theta} = \frac{9}{4}z\left(z-\frac{16}{27}\right),\label{zprime}
\end{equation}
which can be integrated in terms of $\Theta$ time explicitly.

Finally, we could consider the case $\Lambda_{\rm R}<0$.  
The generalized Friedmann constraint equation (\ref{frr}) implies that 
\begin{equation}
\psi^2-\Lambda_{\rm R} \geq 3h^2 \geq 0,
\end{equation}
and we could therefore normalize using
$\sqrt{\psi^2-\Lambda_{\rm R}}$. The resulting dynamical system can be analysed
in a manner similar to that above (see \cite{JMP}).

\subsubsection{Discussion}

The mathematical structure of the dynamics is much richer when there
is a positive RR charge. The dynamics of the system
(\ref{dxdTh})--(\ref{dzdTh}) is of interest from a mathematical point
of view due to the existence of the cyclic behaviour.  The
orbits are future asymptotic to a {\em heteroclinic cycle}, consisting
of the two saddle singular points $S_1$ and $S_2$ and the single
(boundary) orbits in the invariant sets $z=0$ and $1-x^2-z=0$ joining
$S_1$ and $S_2$. The former set corresponds to the zero
$\Lambda_{\rm R}$ solution and
the latter to the solution with constant axion field.  
In a given `cycle' an orbit spends a
long time close to $S_1$ and then moves quickly to $S_2$ shadowing the
orbit in the invariant set $z=0$.  It is then again quasi-stationary
and remains close to the singular point $S_2$ before quickly moving
back to $S_1$ shadowing the orbit in the invariant set $1-x^2-z=0$.
We stress that the motion is {\em not} periodic, and on each
successive cycle a given orbit spends more and more time in the
neighbourhood of the singular points $S_1$ and $S_2$.
The  exact solution corresponding to the singular point $F$ is 
power-law and is defined over the range $-\infty < t <0$,
and  represents a cosmology that collapses monotonically 
to zero volume at $t=0$ in which the   
curvature and coupling are both singular  \cite{JMP}.

The cyclical nature of the orbits  can be physically
understood by reinterpreting the axion field in terms of a 
membrane. Since the axion is 
constant on the surfaces of homogeneity, the field strength 
of the two--form potential must be  directly proportional 
to the volume form of the three--space.  
If the spatial topology of the universe is that of an isotropic 
three--torus, 
the axion field can be formally interpreted as a 
membrane wrapped around this torus \cite{kaloper}. 
All orbits begin at $F$. As the universe collapses, 
the membrane resists being 
squashed into a singular point. This forces the universe 
to bounce into an expansionary phase. The 
cosmological constant then dominates the axion field 
as the latter's energy density decreases. 
The subsequent effect of the cosmological constant can be determined
by viewing Eq. (\ref{generalaction}) in terms of a Brans--Dicke
action, where the coupling parameter between the dilaton and graviton
is given by $\omega =-1$ \cite{bransdicke}; the late--time attractor then
corresponds to a dilaton--vacuum solution.  In effect, therefore, 
the  cosmological constant 
resists the expansion and ultimately 
causes the universe to recollapse and asymptotically 
approach the saddle point $S_1$. 
On the other hand, the collapse causes the 
axion field to become relevant once more and a further bounce 
ensues. The 
process is then repeated with the universe undergoing a series of 
bounces. The orbits move progressively 
closer towards the two saddles, $S_{1,2}$, and spend
increasingly more time near to these points. 
This behaviour is related to the fact that the kinetic energy of the shifted
dilaton field increases monotonically with time, since 
Eq. (\ref{r2}) implies that $\ddot{\varphi} > 0$. It would be interesting
to consider the implications of this behaviour for the pre--big bang
inflationary scenario \cite{pbb}.  

When a modulus field is included $(y \ne 0)$, $F$ still represents the
{\em only} source in the system.  The orbits follow cyclical
trajectories in the neighbourhood of the invariant set $y=0$ and they
spiral outwards monotonically
around the orbit  which
corresponds to the exact solution given by Eqs. (\ref{xfixed})-(\ref{zprime}), with
$dy/d\Theta >0$ from Eq. (\ref{yp}).  After a finite (but arbitrarily large) number of
cycles the kinetic energy of the modulus field becomes more important
until a critical point is reached where it dominates the axion and
cosmological constant. The orbits then asymptote toward the
dilaton--moduli--vacuum solutions corresponding to sources on the line $W$. 

A finite number of modulus fields and shear modes can
be introduced into the model by defining the variable $N^2$ in Eqs. (\ref{4.10}) and
(\ref{y}) via $N^2 = \sum^n_{i=1} N^2_i$.  Orbits with non-trivial modulus field or shear term
 `shadow' the
orbits in the invariant set $y = 0$ at early times and undertake cycles between the
saddles on the singular set $W$
close to $S_1$ and $S_2$. These saddles may be interpreted as
Kasner--like solutions \cite{WE};   the orbits thus experience a finite 
number of cycles in which
the solutions interpolate between different Kasner-like states.  This is 
perhaps reminiscent of the mixmaster behaviour that occurs
in the Bianchi type VIII and IX cosmologies \cite{WE,hobill}.  
Mixmaster oscillations also
occur in less general (i.e., lower-dimensional) Bianchi
models with a magnetic field \cite{LeBlanc1997a} or
Yang-Mills fields \cite{BL}.  It is interesting to note 
in the string context that mixmaster behaviour also
occurs in scalar-tensor theories of gravity in general
and in Brans-Dicke theory in particular \cite{BDT}.
Unlike the mixmaster oscillations, however, the
orbits   eventually spiral away from $y=0$, although
there are orbits that experience a finite but arbitrarily 
large number of oscillations.

Finally, we observe that all of the exact solutions corresponding to
the singular points of the governing autonomous systems of ODEs are
self-similar since in each case the scale factor is a power-law
function of cosmic time.  Therefore, exact self-similar solutions play an important r\^ole in
determining the asymptotic behaviour in string cosmologies.
However, we note that not all of the string solutions are
asymptotically self-similar due to the existence of the heteroclinic
cycle in the RR sector with trivial modulus field.

\vspace{.3in}

\centerline{\bf Acknowledgments}

\vspace{.3in}
I would like to thank Jesus Ib{\'a}{\~n}ez and all of the 
organizers of ERE--99 for inviting me to the
conference and for their warm hospitality.
This work was supported, in part, 
by the Natural Sciences and Engineering Research Council (NSERC)
of Canada.

\end{document}